\documentclass[showpacs,prd]{revtex4}
\usepackage{mathrsfs}
\usepackage{longtable,lscape}
\usepackage{txfonts}
\usepackage{indentfirst}
\usepackage{graphicx,booktabs}
\usepackage{color}
\usepackage{amssymb}
\usepackage{supertabular}
\usepackage{float}

\begin{document}


\title{
Chiral Particle Decay of Heavy-Light Mesons \\[0.2cm]
in a Relativistic Potential Model
}

\author{Takayuki Matsuki$^1$}
\email{matsuki@tokyo-kasei.ac.jp}
\author{Koichi Seo$^2$}
\email{seo@gifu-cwc.ac.jp}

\affiliation{$^1$Tokyo Kasei University, 1-18-1 Kaga, Itabashi, Tokyo 173-8602, Japan \\
$^2$Gifu City Women's College, 7-1 Hito-ichiba Kitamachi, Gifu 501-0192, Japan}

\date{\today}

\begin{abstract}
Partial decay widths of the heavy-light mesons, $D, D_s, B,$ and $B_s$, 
emitting one chiral particle ($\pi$ or $K$) are evaluated 
in the framework of a relativistic potential model. 
Decay amplitudes are calculated by keeping 
the Lorentz invariance as far as possible and use has been made of
the Lorentz-boosted relativistic wave functions of the heavy-light mesons. 
One of predictions of our calculation is very narrow widths of a few keV 
for yet undsicovered 
$B_s(0^+, 1^+)$ mesons corresponding to ${^{2S+1}L_J}={^3P_0}$ and $``{^3P_1}"$ assuming their
masses to be 5617 and 5682 MeV, respectively, 
as calculated in our former paper. 
In the course of our calculation, new sum rules are discovered on the 
decay widths in the limit of $m_Q\to \infty$.  Among these rules, 
$
\Gamma\left(D_{s0}^*(2317)\to D_s+\pi\right)=\Gamma \left(D_{s1}(2460)\to D_s^*+\pi\right)
$
and 
$
\Gamma\left(B_{s0}^*(5615)\to B_s+\pi\right)=\Gamma \left(B_{s1}(5679)\to B_s^*+\pi\right)
$
are predicted to hold with a good accuracy.
\end{abstract}

\pacs{13.25.Gv, 14.40.Pq, 13.75.Lb}
\maketitle

\section{Introduction}
\label{introduction}
\hspace*{\parindent}
We have been trying to explain the mass spectrum of the heavy-light mesons, 
including the famous $D_{s0}(2317)$ and $D_{s1}^*(2460)$, by a relativistic 
potential model with a linear potential and the Coulombic potential. 
In our model besides the current quark mass $m_q$ 
$(m_{u,d}\sim 10 ~{\rm MeV})$ the constant term $b$ is introduced in the scalar
 potential. They are independent parameters in general.  
Howevere, if we restrict our computation up to the first order in $1/m_Q$, 
the current quark mass $m_q$ and the constant $b$ appear only in the 
combination $m_q + b$.  Hence we set $b=0$ and call $m_q$ in place of 
$m_q+b$.  
In this model, we have successfully reproduced many of the experimental 
mass spectrum of 
$D, D_s, B,$ and $B_s$ with a fairy good accuracy including radially excited 
states\cite{mass_spectrum}. 

To confirm the validity of our relativistic potential model,
we have calculated the semileptonic weak form factors (Isgur-Wise functions) for the process
$\bar B\to D^{(*)}\ell\bar\nu$ in Ref. \cite{MS1} to obtain reasonable results compared
with the experiments. 
In the present paper \cite{MS2}, 
we calculate the decay processes of the heavy-light mesons with one chiral particle emission.  
Although the decay rates of these processes have
been examined by Goity and Roberts~\cite{GR} and by Di Pierro and Eichten
~\cite{PE} in a relativistic potential model, we would like to see the effects of using our relativistic wave functions as well as Lorentz covariant amplitudes that
have been neglected in their papers.
The authors of Refs. \cite{GR} and \cite{PE} have used wave functions of the heavy-light mesons in the rest frame to calculate 
transition amplitudes.  That is, they have neglected not only the relativistic effects of the
wave functions but also the recoil effects of the heavy-light mesons.
We claim that this recoil effect is essential to calculate the relativistic
transition amplitude because the plane wave function of the emitted particle is replaced by another phase factor, which involves the heavy quark mass and the velocity of the heavy-light meson.
The plane wave, which, for instance, is used in a hydrogen atom, is inserted without any criticism, which may lead to the erroneous conclusion.
In the course of our calculation, a couple of new sum rules on the decay 
widths are discovered, which hold among the different decay processes in the 
heavy quark limit, i.e., $m_Q\to\infty$.  The breaking of the sum rules is 
of the order of $(M_i-M_f)/m_Q$, which is at most 10 $\%$ or so with the initial
and final heavy-light meson masses, $M_i$ and $M_f$.

In Sec. \ref{lorentz}, the wave function of the heavy-light meson in the moving frame is 
related to the one in the rest frame by regarding the heavy quark to be at 
rest in the heavy-light meson.  
Then we derive the basic formula to express the transition amplitude 
in terms of the wave functions of the heavy-light meson in the rest frame.  
The plane wave of the emitted particle moving in the $z$-direction, $\exp\left(-ikz\right)$,
used in other 
works~\cite{GR} and~\cite{PE} is found to be replaced by another phase factor, 
$\exp\left(-2im_QVz\right)$, with $V$ being the heavy-light meson velocity in the Breit frame.
In the limit of $m_Q\to\infty$, the heavy quark mass $m_Q$
coincides with the heavy-light meson mass and these two phase factors become 
equal to each other.

Interaction between light quarks and chiral particles ($\pi$ or $K$) is assumed 
{\it a la} Georgi-Manohar~\cite{GM}.  The matrix elements of axial-vector currents 
are dissolved into several form factors, which are expressed in terms of the 
radial wave functions.

In Sec. \ref{sec3}, we numerically evaluate the matrix elements of axial-vector 
currents.  Except for the axial-vector coupling constant, We adopt the parameters of
the model which have been determined in the previous paper to fit with the experimental 
mass spectra.  
The axial-vector coupling constant is adjusted so that the 
theoretical decay widths of $D^{*+}(2010)$ coincide with the experimental values,
%
whose derivation is still contentious and is discussed in Sec. \ref{sec5}.
%
We predict the decay widths of the various decay modes, most of which have 
not yet been experimentally observed.  The total widths of some heavy-light 
mesons are experimentally known, which are compared with our theoretical 
results.  
Because $D_s\left(1^-\right)$, $D_s\left(0^+\right)$, and $D_s\left(1^+\right)$ are not
allowed kinematically to emit 
one $K$ meson, assuming the mixing of $\pi^0$ and $\eta$ we have the 
narrow decay widths of these particles.  Likewise, we predict very narrow 
decay widths of $B_s\left(0^+\right)$ and $B_s\left(1^+\right)$, which are not yet observed  
and hence we have used our theoretical 
values of their masses, 5615 MeV and 5679 MeV, respectively.

In Sec. \ref{sec4}, new sum rules on the decay widths are presented in the limit of $m_Q\to\infty$.  In this limit the spin of the heavy quark is decoupled from 
the orbital angular momentum and the spin of the light quark, and the heavy 
quark is no more than a spectator in the decay process.  The radial wave 
functions of $J^P$ and $(J+1)^P$ coinside with each other and their masses 
become degenerate.  We call that these degenerate particles belong to a spin
multiplet.  The decay widths of particles in the same multiplet become equal 
if we sum over the final states in another multiplet.

Sec. \ref{sec5} is devoted to the conclusions and discussion.  
In Appendix A, the matrix elements of the axial-vector currents are described 
by the polarization vectors/tensors in the rest frame of each heavy-light meson.
In Appendix B, the properties of the polarization vectors are given as well as 
the explicit forms of angular-spin part of the wave function.
In Appendix C, the transition amplitudes are expressed in terms of the radial 
wave functions and the spherical Bessel function, $f(x)\equiv \sin{x}/x$.

\section{Lorentz Invariant Evaluation of Transition Amplitudes}
\label{lorentz}

\hspace*{\parindent}
The wave function of a heavy-light meson with a finite momentum is defined as
\begin{eqnarray}
<0\,|\,q^c_i({\vec x},t)\,Q_j({\vec y},t)\,|\,P>
&=&<0\,|\,q^c_i({\vec x}-{\vec X}_{\xi},0)\,
Q_j({\vec y}-{\vec X}_{\xi},0)\,|\,P>
e^{-iP\cdot X_{\xi}}\nonumber\\
&=&\psi^{(\xi)}_{ij}({\vec x}-{\vec y};P)\,e^{-iP\cdot X_{\xi}},
\end{eqnarray}
where $X_{\xi}~(=\xi x+(1-\xi)y)$ denotes the position of the heavy-light
 meson and $\xi$ is a free parameter.  If we set $\xi=0$ or $1$, then $X_\xi$
 coincides with the position of the heavy quark or light quark, respectively.

We assume chiral interaction of pseudoscalar mesons ($\pi$ and $K$) with
 light quarks.  In the present paper we only compute one pseudoscalar particle
 emission from heavy-light mesons, and then the relevant interaction
 Lagrangian is as follows:
\begin{equation}
\mathcal{L}_{int}=\frac{g}{\sqrt{2}f_\pi}
\bar{q}_i\gamma_\mu\gamma_5q_j\partial^\mu\phi_{ij}, \label{Lint}
\end{equation}
where $g$ is a dimensionless coupling constant and $f_\pi$ is the 
pion decay constant.  Here the flavor SU(3) symmetry
 is assumed and 
$\phi_{ij}$'s represent the octet meson fields, that is,
\begin{equation}
\left(\phi_{ij}\right)=\sqrt{2}
\left(\begin{array}{ccc}
\frac{1}{\sqrt{2}}\phi_3+\frac{1}{\sqrt{6}}\phi_8
& \pi^+ & K^+
\\
\pi^- &
-\frac{1}{\sqrt{2}}\phi_3+\frac{1}{\sqrt{6}}\phi_8
& K^0
\\
K^- & \overline{K^0} &
-\frac{2}{\sqrt{6}}\phi_8
\end{array}\right). \label{eq:octet}
\end{equation}
The mixing of $\pi^0$ and $\eta$ is taken into account 
with a small parameter $\epsilon$ as follows:
\begin{equation}\left(
\begin{array}{c}
\pi^0 \\ \eta
\end{array}\right)
=\frac{1}{\sqrt{1+\epsilon^2}}\left(\begin{array}{cc}
1 & \epsilon\\
-\epsilon & 1
\end{array}\right)
\left(\begin{array}{c}\phi_3 \\ \phi_8\end{array}\right).
\end{equation}

The transition amplitude for a heavy-light meson with a momentum $P$ into 
another heavy-light
meson with $P'$ and one pseudoscalar particle with $k$ is transformed
by inserting the number operator of the heavy quark and by neglecting the 
sea quark effects as follows:
\begin{eqnarray}
\mathcal{M}_{P\to P'}&=&\int d^4x \,<P',k\,|\,\mathcal{L}_{int}(x)\,|\,P>\nonumber\\
&=&\int d^4x  \,<P',k\,|\,\mathcal{L}_{int}(x)
\int d^3y\,Q^\dagger_\ell(y)Q_\ell(y)\,|\,P>\nonumber\\
&\approx&\int d^4x\int d^3y\,O_{ij}
<P'\,|Q^\dagger_\ell(y)q^{c\dagger}_i(x)|\,0>
<0\,|q^c_j(x)Q_\ell(y)|\,P>e^{ik\cdot x}\nonumber\\
&=&\int d^4x\int d^3y\,
\text{tr}\left[\psi'^{(\xi)\dagger}_{i\ell}(\vec{x}-\vec{y};P')
\,O_{ij}\,\psi^{(\xi)}_{j\ell}(\vec{x}-\vec{y};P)\right]\;
e^{ik\cdot x-i(P-P')\cdot X_\xi}\nonumber\\
&=&(2\pi)^4\delta^4(P-P'-k)\int d^3z\;
\text{tr}\left[\psi'^{(0)\dagger}_{i\ell}({\vec z};P')
\,O_{ij}\,\psi^{(0)}_{j\ell}({\vec z};P)\right]\;
e^{-i{\vec k}\cdot{\vec z}}\label{eq:basic0}\\ 
&=&(2\pi)^4\delta^4(P-P'-k)\int d^3z\;\text{tr}\left[
\psi'^{(1)\dagger}_{i\ell}({\vec z};P')
\,O_{ij}\,\psi^{(1)}_{j\ell}({\vec z};P)\right]\label{eq:basic1}
\end{eqnarray}

In Ref.~\cite{MS1} the following approximate relation has been obtained between the wave function with a finite momentum ($P_3=\gamma MV$) and the one at rest,
\begin{equation}
\psi^{(0)}_{\alpha\beta}({\vec x};P)\approx
G_{\alpha\gamma}G_{\beta\delta}\psi_{\gamma\delta}(x_\perp,\gamma z;M)\,
e^{i(M-m_Q)V\gamma z},\label{eq:boost1}
\end{equation}
and the following relation is derived likewise,
\begin{equation}
\psi^{(1)}_{\alpha\beta}({\vec x};P)\approx
G_{\alpha\gamma}G_{\beta\delta}\psi_{\gamma\delta}(x_\perp,\gamma z;M)\,
e^{-im_QV\gamma z},\label{eq:boost2}
\end{equation}
where $G$ is the boost operator from the rest frame to the moving frame
and $\psi(x; M)$ is the wave function in the rest frame.
The Lorentz transformation from the rest
frame to the moving frame produces the time difference between two constituent
quarks, and this time difference is compensated by the free propagation of 
the heavy quark, which gives rise to the phase factors in 
Eqs.(\ref{eq:boost1}) and (\ref{eq:boost2}).

If we evaluate the transition amplitude in the Breit frame, 
where the initial heavy-light meson is moving
to the z direction with a velocity $V$
 and the final heavy-light meson is moving in the opposite direction
with the same velocity $V$, 
and insert Eqs.(\ref{eq:boost1}) and (\ref{eq:boost2}) 
into Eqs.(\ref{eq:basic0}) and (\ref{eq:basic1}), 
we obtain the following unique expression,
\begin{eqnarray}
\mathcal{M}_{P\to P'}&\approx&
(2\pi)^4\delta^4(P-P'-k)\nonumber\\
&&\qquad\times\int d^3x\;\text{tr}\left[(G^{-1})^T
\psi'^\dagger_{i\ell}(x_\perp,\gamma z;M')G^{-1}
\,O_{ij}G\,\psi_{j\ell}(x_\perp,\gamma z;M)G^T\right]e^{-2im_Q\gamma Vz}\nonumber\\
&=&(2\pi)^4\delta^4(P-P'-k)\nonumber\\
&&\qquad\times\gamma^{-1}\int d^3x\;\text{tr}\left[(G^{-1})^T
\psi'^\dagger_{i\ell}(\vec{x};M')G^{-1}\,O_{ij}G\,
\psi_{j\ell}(\vec{x};M)G^T\right]e^{-2im_QVz}.\label{eq:basic_amp}
\end{eqnarray}
Here two points should be noted. 
First this is independent of $\xi$ as it should be.
Next the the phase factor $e^{-2im_QVz}$ has appeared in place of 
the wave function of the emitted pseudoscalar particle, $e^{-ikz}$,
that is absorbed into the delta function to preserve the four-momentum conservation
in either Eq.(\ref{eq:basic0}) or (\ref{eq:basic1}).
This is the result of taking into account of the heavy-meson's recoil
effect.

In general the matrix elements of the axial-vector current between the various 
spin states of the heavy-light mesons have the following tensor structures,
\begin{eqnarray}
\frac{<0^-|j_{5\mu}|1^->}{\sqrt{M_2M_1}}&=&
(1+\omega)\varepsilon_{1\mu}\xi^{(k)}_{A1}(\omega)\nonumber\\
&&\hskip -10pt
+(\varepsilon_1\cdot v_2)\left\{
(v_1+v_2)_\mu\xi^{(k)}_{A2}(\omega)+
(v_1-v_2)_\mu\xi^{(k)}_{A3}(\omega)\right\}\quad(k=1,10), \label{OneSD}
\\
\frac{<0^-|j_{5\mu}|0^+>}{i\sqrt{M_2M_1}}&=&
(v_1+v_2)_\mu\xi^{(2)}_{A1}(\omega)
+(v_1-v_2)_\mu\xi^{(2)}_{A2}(\omega),
\\
\frac{<1^-|j_{5\mu}|0^+>}{\sqrt{M_2M_1}}&=&
\epsilon_{\mu\nu\rho\sigma}v_1^{\nu} v_2^{\rho}
\varepsilon_2^{*\sigma}\xi^{(3)}_A(\omega),
\\
\frac{<0^-|j_{5\mu}|1^+>}{\sqrt{M_2M_1}}  &=&
\epsilon_{\mu\nu\rho\sigma}v_1^{\nu} v_2^{\rho}
\varepsilon_1^{\sigma}\xi^{(k)}_A(\omega)\quad(k=4,5), \label{OneP1}
\\
\frac{<1^-|j_{5\mu}|1^+>}{i\sqrt{M_2M_1}}&=&
(\varepsilon^*_2\cdot\varepsilon_1)(v_1+v_2)_\mu\xi^{(k)}_{A1}(\omega)
+(\varepsilon^*_2\cdot\varepsilon_1)(v_1-v_2)_\mu\xi^{(k)}_{A2}(\omega)
\nonumber\\
&& \qquad\quad\quad
+(\varepsilon^*_2\cdot v_1)\varepsilon_{1\mu}\xi^{(k)}_{A3}(\omega)
+(\varepsilon_1\cdot v_2)\varepsilon^*_{2\mu}\xi^{(k)}_{A4}(\omega)\quad(k=6,7), \label{OneP2}
\\
\frac{<0^-|j_{5\mu}|2^+>}{i\sqrt{M_2M_1}} &=&
\varepsilon_{1\mu\nu}v_2^{\nu}\xi^{(8)}_{A1}(\omega)+
\varepsilon_{1\alpha\beta}v_2^{\alpha}v_2^{\beta}
\left\{v_{1\mu}\xi^{(8)}_{A2}(\omega)+
v_{2\mu}\xi^{(8)}_{A3}(\omega)\right\},
\\
\frac{<1^-|j_{5\mu}|2^+>}{\sqrt{M_2M_1}} &=&
\epsilon_{\mu\nu\rho\sigma}\biggl[
\varepsilon_1^{\nu\alpha}v_{2\alpha}\varepsilon^{*\rho}_2\left\{
(v_1+v_2)^\sigma\xi^{(9)}_{A1}(\omega)+
(v_1-v_2)^\sigma\xi^{(9)}_{A2}(\omega)\right\}
\nonumber\\
&&\hskip -20pt
+v^\rho_1v^\sigma_2\left\{
\varepsilon^{\nu\alpha}_1\varepsilon^*_{2\alpha}\xi^{(9)}_{A3}(\omega)
+\varepsilon_1^{\nu\alpha}v_{2\alpha}(\varepsilon^*_2\cdot v_1)
\xi^{(9)}_{A4}(\omega)
+\varepsilon_{1\alpha\beta}v_2^\alpha v_2^\beta
\varepsilon^{*\nu}_2\xi^{(9)}_{A5}(\omega)
\right\}\biggr]\nonumber\\
&&
+\epsilon_{\alpha\beta\gamma\delta}\varepsilon_1^{\alpha\alpha'}v_{2\alpha'}
\varepsilon^{*\beta}_2v^\gamma_1v^\delta_2
\left\{(v_1+v_2)_\mu\xi^{(9)}_{A6}(\omega)
+(v_1-v_2)_\mu\xi^{(9)}_{A7}(\omega)\right\},\\
\frac{<1^-|j_{5\mu}|1^->}{i\sqrt{M_2M_1}} &=&
\epsilon_{\mu\nu\rho\sigma}\biggl[
\varepsilon_1^{\nu}\varepsilon^{*\rho}_2\left\{
(v_1+v_2)^\sigma\xi^{(11)}_{A1}(\omega)+
(v_1-v_2)^\sigma\xi^{(11)}_{A2}(\omega)\right\}
\nonumber\\
&&\quad
+v^\rho_1v^\sigma_2\left\{
\varepsilon^{\nu}_1(\varepsilon^*_{2}\cdot v_1)\xi^{(11)}_{A3}(\omega)
+\varepsilon_2^{*\nu}(\varepsilon_1\cdot v_2)\xi^{(11)}_{A4}(\omega)\right\}
\biggr]\nonumber\\
&&\hskip 10pt
+\epsilon_{\alpha\beta\gamma\delta}\varepsilon_1^{\alpha}
\varepsilon^{*\beta}_2v^\gamma_1v^\delta_2
\left\{(v_1+v_2)_\mu\xi^{(11)}_{A5}(\omega)
+(v_1-v_2)_\mu\xi^{(11)}_{A6}(\omega)\right\},
\end{eqnarray}
where the initial and final quantities have subindices 1 and 2, respectively, 
$\varepsilon_{i\mu}$ and $\varepsilon_{i\mu\nu}$ are polarization vector and tensor,
$v_1$ and $v_2$
are velocity vectors defined by $P_1=M_1v_1$ and $P_2=M_2v_2$, 
which in the Breit frame have components, 
$v_1^0=v_2^0=\gamma,\quad v_1^3=-v_2^3=\gamma V$, and 
$\omega=v_1\cdot v_2=P_1\cdot P_2/(M_1M_2)$, which in the Breit frame is related to $V$ by
\begin{equation}
\omega=\gamma^2(1+V^2)=\frac{1+V^2}{1-V^2}.
\end{equation}
In our model the $1^+$ mass-eigenstates appearing in Eqs.~(\ref{OneP1}) and (\ref{OneP2})
are characterized by the quantum number of $j=L+s_q$, and $^3P_1$ and $^1P_1$ dominant
states are denoted by $"^3P_1"$ and $"^1P_1"$, respectively. This is the reason why
we have two kinds of form factors in these equations, i.e., $k=4, 5$ and $k=6, 7$.
There are also two $1^-$ states with $k=1,10$ in Eq.~(\ref{OneSD}) corresponding to
$^3S_1$ and $^3D_1$ states, respectively.
Each componet of the matrix elements of the axial-vector current in the Breit 
frame is written down with the polarization vectors/tensors 
in each rest frame of the initial and final mesons in Appendix A.

In order to obtain explicit forms of the form factors $\xi_A$'s, we need to calculate
the matrix elements by inserting the explicit forms of the wave functions.
According to Eq.(\ref{eq:basic_amp}) the matrix elements of
the axial vector current are evaluated in terms of the wave functions 
in the rest frame of the heavy-light mesons, $\psi_i(\vec r,M)$, as follows:
\begin{eqnarray}
\frac{<\psi_2|j_5{}^\mu|\psi_1>}{2\sqrt{M_1M_2}} &\approx&
\gamma^{-1}\int d^3x\,\frac{1}{4\pi r^2}
\frac{1}{2}\text{tr}\left[y^*_2\left(
u_2,i({\vec n}\cdot{\vec \sigma})v_2\right)G^{-1}
(\rho_1,\sigma_1,\sigma_2,\sigma_3)G 
\biggl(\begin{array}{c}u_1\ \\
 -i({\vec n}\cdot{\vec \sigma})v_1\end{array}\biggr)
y_1\right]e^{iqz}, \\
  \psi_i(\vec r;M) &=&  \left(\begin{array}{c}u_i(r)\ \\
  -i({\vec n}\cdot{\vec \sigma})v_i(r)\end{array}\right) y_i, 
\end{eqnarray}
where $q=-2m_QV$, $\sigma_i\equiv\sigma_i\otimes 1_{2\times 2}$,
and $\vec n={\vec r}/{r}$.
By substitution of the explicit expression for $G$, 
the matrix elements of each component of the axial vector current 
are obtained as follows:
\begin{eqnarray}
\frac{<\psi_2|j_5{}^0|\psi_1>}{2\sqrt{M_1M_2}} &\approx&
-i\gamma^{-1}\int d^3x \,\frac{1}{4\pi r^2}
(u_2v_1-v_2u_1)\frac{1}{2}\text{tr}\left[y^*_2
({\vec n}\cdot{\vec \sigma})
y_1\right]\,e^{iqz},
\\
\frac{<\psi_2|j_5{}^3|\psi_1>}{2\sqrt{M_1M_2}} &\approx&
\gamma^{-1}\int d^3x \,\frac{1}{4\pi r^2}
u_2u_1\frac{1}{2}\text{tr}\left[y^*_2\sigma_3
y_1\right]\,e^{iqz}\nonumber\\
&&+\gamma^{-1}\int d^3x \,\frac{1}{4\pi r^2}
v_2v_1\frac{1}{2}\text{tr}\left[y^*_2
({\vec n}\cdot{\vec \sigma})\sigma_3({\vec n}\cdot{\vec \sigma})
y_1\right]\,e^{iqz},
\\
\frac{<\psi_2|j_5{}^i|\psi_1>}{2\sqrt{M_1M_2}} &\approx&
\int d^3x \,\frac{1}{4\pi r^2}
u_2u_1\frac{1}{2}\text{tr}\left[y^*_2\sigma_i
y_1\right]\,e^{iqz}\nonumber\\
&&+\int d^3x \,\frac{1}{4\pi r^2}
v_2v_1\frac{1}{2}\text{tr}\left[y^*_2
({\vec n}\cdot{\vec \sigma})\sigma_i({\vec n}\cdot{\vec \sigma})
y_1\right]\,e^{iqz}\nonumber\\
&&+\int d^3x \,\frac{1}{4\pi r^2}
\epsilon_{3ij}V
\frac{1}{2}\text{tr}\biggl[y^*_2\left\{v_2u_1
({\vec n}\cdot{\vec \sigma})\sigma_j-
u_2v_1\sigma_j({\vec n}\cdot{\vec \sigma})\right\}
y_1\biggr]\,e^{iqz}\quad (i=1,2).
\end{eqnarray}
The angular-spin part of wave functions, $y_i(\Omega)$, are given in Appendix B.

The form factors
 $\xi_A$'s are expressed in terms of the radial
wave functions $u$'s and $v$'s in Appendix C, where we have adopted only 
the terms up to the first order in $V$.  The formula for the decay width,
\begin{equation}
\Gamma=\frac{k_R}{8\pi M_1^2}\frac{1}{2j_1+1}
\sum_{pol}|<P_2,k|\mathcal{L}_{int}(0)|P_1>|^2,
\end{equation}
is also evaluated in terms of $\xi_A$'s in Appendix C, where $k_R$ is the 
momentum of the chiral particle in the rest frame of the parent heavy-light meson.

If the mass difference of the parent heavy-light meson with strangeness 
and daughter non-strange meson is smaller than 
the $K$-on mass, then the parent heavy-light meson with strangeness decays 
into a heavy-light meson with strangeness and a neutral pion through 
the $\pi^0-\eta$ mixing.  
The mixing parameter $\epsilon$ has been related to the current quark masses 
due to Gasser and Leutwyler ~\cite{gass} as follows:
\begin{equation}
\epsilon=\frac{\sqrt{3}}{4}\frac{m_d-m_u}{m_s-(m_u+m_d)/2}
=1.00\times10^{-2}
\end{equation}
As a result, the formula of the decay width is multiplied by the factor,
 $\epsilon^2 ~(=1.00\times 10^{-4})$, in addition to the factor $2/3$ 
(the square of the coefficient of $\phi_8$ in the octet meson matrix 
(\ref{eq:octet})).

\section{Numerical Results}\label{sec3}
\hspace*{\parindent}
In the leading order of $1/m_Q$ expansion, the radial wave functions 
$u(r)$ and $v(r)$ are determined by the following Dirac equation,
\begin{equation}\left(
\begin{array}{cc}
m_q+S+V & -\frac{\partial}{\partial r}+\frac{k}{r}\\
\frac{\partial}{\partial r}+\frac{k}{r} & -m_q-S+V
\end{array}\right)
\left(\begin{array}{c}
u(r)\\
v(r)
\end{array}\right)=
E\left(\begin{array}{c}
u(r)\\
v(r)
\end{array}\right),
\end{equation}
where $S$ and $V$ are confining scalar and Coulombic 
potentials parametrized by 
$S(r)=b+\frac{r}{a^2}$ and $V(r)=\frac{4\alpha_s}{3r}$.  $k$ is the eigenvalue of 
the operator $-\beta_q({\vec L}\cdot{\vec \sigma_q}+1)$.  The states with the 
total angular momentum $j$ and $j+1$ have the same value of $k$, and they have the same 
radial wave function and the same eigenvalue $E$.  For example, $S$-wave 
states $0^-$ and $1^-$ with $k=-1$ are degenerate.  Taking into account  
 the asymptotic behaviors at infinity and at the origin, $u(r)$ and $v(r)$ are 
approximated by the following forms,
\begin{eqnarray}
u(r)&=&r^\gamma \exp\{-r^2/a^2-(b+m_q)r\}\sum_{i=0}^7 a^{(u)}_ir^i,\\
v(r)&=&r^\gamma \exp\{-r^2/a^2-(b+m_q)r\}\sum_{i=0}^7 a^{(v)}_ir^i,
\end{eqnarray}
where 
$\gamma=\sqrt{k^2-(\frac{4\alpha_s}{3})^2}$.
The values of the parameters used in the numerical calculation are listed in
 Table~\ref{tab:param}, which are determined so as to reproduce the mass spectra of the heavy-light mesons in the order of $1/m_Q$.
(See Ref.~\cite{mass_spectrum}.)     In the present calculation the constant
 $b$ and the light quark mass appear only in the linear combination, and the 
values of this combination are given in the table.

\begin{table}[H]
\caption{Values of the parameters}
\label{tab:param}
\begin{center}
\begin{tabular}{|c|c|c|c|c|c|c|} \hline
$\alpha_s$ & 
\begin{tabular}{c}$a$\\(GeV$^{-1}$)\end{tabular} &
\begin{tabular}{c}$b+m_u(=m_d)$\\(MeV/$c^2$)\end{tabular} &
\begin{tabular}{c}$b+m_s$\\(MeV/$c^2$)\end{tabular} &
\begin{tabular}{c}$m_c$\\(MeV/$c^2$)\end{tabular} &
\begin{tabular}{c}$m_b$\\(MeV/$c^2$)\end{tabular} &
\begin{tabular}{c}$g^2$\end{tabular} \\
\hline
\begin{tabular}{c}0.259($D$)\\0.392($B$)\end{tabular}&
1.937&86&168&1023&4634 & 0.608 \\\hline
\end{tabular}
\end{center}
\end{table}
Here different values of the strong coupling $\alpha_s$ are used for $D/D_s$ and
$B/B_s$ mesons.
This parameter set is not a unique solution, but 
the results are not so changed even if we take another set of values given 
in Ref.~\cite{MS1}.
The axial-vector coupling constant $g$ is determined so that the calculated 
decay widths of $D^{*+}(2010)$ agree with the experimental values.  The value of 
$g^2$ is taken as $0.608$ with a $27\%$ uncertainty.  
The errors of prediction 
are caused by the uncertainty of our parameters besides $g^2$, and only the 
central values of the calculated widths are shown in Table 2, 3, 4 and 5.  
$k_R$ is the momentum of the emitted chiral particle in the rest frame of
 the parent heavy-light meson.  The observed widths in our tables are taken 
from Ref.~\cite{PDF}, and experimental values depending on model assumptions 
are omitted.


\begin{center}
\label{tab:tab_res1}
\tablecaption{Numerical evaluation of the decay widths of excited $D$ mesons}
\tablefirsthead{\hline
Initial State & Final & $k_R$ & $\Gamma_{th}$ 
& $\Gamma_{exp}$ \\
$\left({}^{2S+1}L_J\right)$ & State & (MeV/c) & (MeV)
& (MeV) \\ \hline
}
\tablehead{\hline
Initial State & Final & $k_R$ & $\Gamma_{th}$ 
& $\Gamma_{exp}$ \\
$\left({}^{2S+1}L_J\right)$ & State & (MeV/c) & (MeV) 
& (MeV) \\ \hline
}
\begin{supertabular}{|c|c|c|c|c|}
$D^{*0}$ & $D^\pm\pi^\mp$ & - & 
 - & not allowed \\ 
$({}^3S_1)$ & $D^0\pi^0$ & 43.1 & 
 0.042 & $<1.3$\\ \hline
$D^{*\pm}$ & $D^0\pi^\pm$ & 39.5 & 
 Input & $0.065\pm0.018$ \\ 
$({}^3S_1)$ & $D^\pm\pi^0$ & 38.2 & 
 Input & $0.029\pm0.08$ \\ \hline
\hline
 & $D^\pm\pi^\mp$ & 414.2 & 
 $0.99\times10^2$ & \\ 
$D_0^{*}(2400)^0$ & $D^0\pi^0$ & 419.4 & 
 $0.50\times10^2$ & \\ 
$({}^3P_0)$ & $D\pi$(sum) & & $2.5\times10^2$ & \\ 
& all & & & $261\pm50$\\
\hline
 & $D^0\pi^\pm$ & 461.3 & 
 $1.2\times10^2$ & \\ 
$D_0^{*}(2400)^\pm$ & $D^\pm\pi^0$ & 458.5 & 
 $0.61\times10^2$ & \\ 
$({}^3P_0)$ & $D\pi$(sum) & & $1.8\times10^2$ & \\ 
& all & & & $283\pm24\pm34$\\
\hline
$D_0^{*}(2400)$ & $D^*\pi$ &  & 
 0 & not seen\\ \hline
\hline
$D_1(2430)$ & $D\pi$ &  & 
 0 & \\ \hline
& $D^{*\pm}\pi^\mp$ & 358.7 & 
 $0.76\times10^2$ & \\ 
$D_1(2430)^0$ & $D^{*0}\pi^0$ & 363.1 & 
 $0.39\times10^2$ & \\ 
$("{}^3P_1")$ & $D^*\pi$(sum) & & $1.9\times10^2$ & \\ 
& all & & & $384\begin{array}{c}+107\\-75\end{array}\pm74$
\\ \hline
\hline
$D_1(2420)$ & $D\pi$ &  & 
 0 & not seen\\ \hline
& $D^{*\pm}\pi^\mp$ & 354.5 & 
 2.7 & \\ 
$D_1(2420)^0$ & $D^{*0}\pi^0$ & 358.9 & 
 1.5 & \\ 
$("{}^1P_1")$ & $D^*\pi$(sum) & & 6.9 & \\ 
& all & & & $20.4\pm1.7$\\ \hline
& $D^{*0}\pi^\pm$ & 358.4 & 
 2.9 & \\ 
$D_1(2420)^\pm$ & $D^{*\pm}\pi^0$ & 357.0 & 
 1.4 & \\ 
$("{}^1P_1")$ & $D^*\pi$(sum) & & 4.3 & \\ 
& all & & & $25\pm6$\\ \hline
\hline
 & $D^{\pm}\pi^\mp$ & 505.5 & 
 6.8 & \\ 
$D_2^{*}(2460)^0$ & $D^{0}\pi^0$ & 510.2 & 
 3.6 & \\ 
$({}^3P_2)$ & $D^{*\pm}\pi^\mp$ & 389.2 & 
 2.6 & \\ 
 & $D^{*0}\pi^0$ & 393.4 & 
 1.4 & \\  
& $D\pi+D^*\pi$(sum) & & 24 & \\ 
& all & & & $43\pm4$\\
\hline
 & $D^{0}\pi^\pm$ & 508.4 & 
 7.1 & \\ 
$D_2^{*}(2460)^\pm$ & $D^{\pm}\pi^0$ & 505.6 & 
 3.4 & \\ 
$({}^3P_2)$ & $D^{*0}\pi^\pm$ & 391.2 & 
 2.7 & \\ 
 & $D^{*\pm}\pi^0$ & 389.7 & 
 1.3 & \\ 
& $D\pi+D^*\pi$(sum) & & 15 & \\ 
& all & & & $37\pm6$\\ \hline
\pagebreak
 & $D^{\pm}\pi^\mp$ & 739.6 & 
 4.8 & \\ 
$D(2760)^0$ & $D^{0}\pi^0$ & 743.6 & 
 2.4 & \\ 
$({}^3D_1)$ & $D^{*\pm}\pi^\mp$ & 638.9 & 
 0.79 & \\ 
 & $D^{*0}\pi^0$ & 642.1 & 
 0.41 & \\ 
& $D\pi+D^*\pi$(sum) & & 14 & \\ 
& all & & & $60.9\pm5.1$\\ \hline
 & $D^{0}\pi^\pm$ & 742.9 & 
 4.9 & \\ 
$D(2760)^\pm$ & $D^{\pm}\pi^0$ & 740.3 & 
 2.4 & \\ 
$({}^3D_1)$ & $D^{*0}\pi^\pm$ & 641.3 & 
 0.81 & \\ 
 & $D^{*\pm}\pi^0$ & 639.6 & 
 0.39 & \\ 
& $D\pi+D^*\pi$(sum) & & 8.5 & \\ 
& all & & & no exp\\ \hline
\end{supertabular}
\end{center}

The narrow width of $D^*(1^-)$ is due to the smallness of the Q value.  The 
width of $D_1(2420)$ is smaller than that of $D_1(2430)$ by an order of 
magnitude.  The values of $k$ of the former and the latter are assumed to be 
$-2$ and $1$, respectively, and their angular momenta 
neglecting the heavy quark spin denoted as $j_\ell$ are 
$\frac{3}{2}$ and $\frac{1}{2}$,respectively.  
By the parity and the angular momentum 
conservation, $1^+$ is able to decay into $1^-$ and one pseudoscalar 
particle with an orbital angular momentum of $L=0\hbox{ or }2$.  Since 
$j_\ell$ of the daughter meson $D^*(1^-)$ is $\frac{1}{2}$, 
$D_1(2420)(k=-2)$ emits 
a pseudoscalar particle with $L=2$ only, while $D_1(2430)(k=1)$ emits one 
with $L=0$.  This is the widely accepted explanation of the difference of the 
decay widths of $D_1(2420)$ and $D_1(2430)$.  As for the time componet of the 
axial vector current, this argument is true and the ratio of the matrix 
elements is almost one-thousandth.  As a matter of fact, the space component 
parallel to the momentum of the emitted particle in the Breit frame is not so 
suppresed, and the ratio of the widths amounts to 10.

\begin{center}
\label{tab:tab_res2}
\tablecaption{Numerical evaluation of the decay widths of excited $D_s$ mesons}
\tablehead{\hline
Initial State & Final & $k_R$ & $\Gamma_{th}$ 
& $\Gamma_{exp}$ \\
$\left({}^{2S+1}L_J\right)$ & State & (MeV/$c$) & (MeV) 
& (MeV) \\ \hline
}
\begin{supertabular}{|c|c|c|c|c|}
$D_s^{*\pm}({}^3S_1)$ & $D_s^\pm\pi^0$ & 47.8 & 
$8.0\times10^{-6}$ & $<0.11$ \\ \hline
\hline
$D_{s0}^{*}(2317)^\pm$ & $D_s^\pm\pi^0$ & 297.7 & 
$3.8\times10^{-3}$ & $<3.8$\\ 
$({}^3P_0)$ & $D_s^{*\pm}\pi^0$ & 148.0 & 
0 & \\ \hline
\hline
$D_{s1}(2460)^\pm$ & $D_s^{\pm}\pi^0$ & 424.8 & 
0 & \\ 
$("{}^3P_1")$ & $D_s^{*\pm}\pi^0$ & 297.3 & 
$3.9\times10^{-3}$ & $<1.7$\\ \hline
\hline
 & $D^{0}K^\pm$ & 390.4 & 
0 & not seen\\ 
$D_{s1}(2536)^\pm$ & $D^{\pm}K^0$ & 385.6 & 
0 & not seen\\ 
$("{}^1P_1")$ & $D^{*0}K^\pm$ & 165.3 & 
0.066 & \\ 
 & $D^{*\pm}K^0$ & 159.4 & 
0.055 & \\ 
& $D^*K$(sum) & & 0.12 & \\ 
& all & & & $<2.3$\\ \hline
\hline
 & $D^0K^\pm$ & 433.9 & 
3.4 & \\ 
$D_{s2}(2573)^\pm$ & $D^\pm K^0$ & 429.4 & 
3.3 & \\ 
$({}^3P_2)$ & $D^{*0}K^\pm$ & 242.6 & 
0.27 & \\ 
 & $D^{*\pm}K^0$ & 238.3 & 
0.24 & \\ 
& $DK+D^*K$(sum) & & 7.2 & \\ 
& all & & & $20\pm5$\\ \hline
\hline
 & $D^0K^\pm$ & 673.2 & 
6.6 & \\ 
$D_{s}(2818)^\pm$ & $D^\pm K^0$ & 669.8 & 
6.4 & \\ 
$({}^3D_1)$ & $D^{*0}K^\pm$ & 547.3 & 
1.1 & \\ 
not observed & $D^{*\pm}K^0$ & 544.8 & 
1.0 & \\ 
& $DK+D^*K$(sum) & & 15 & \\ \hline
\end{supertabular}
\end{center}

$D_s^*$, $D_{s0}^*(2317)$, and $D_{s1}(2460)$ are kinematically forbidden to 
emit a strange particle, and their decays proceed through the $\pi^0-\eta$ 
mixing.  The mixing parameter $\epsilon$ is assumed to be $0.01$, and these 
decay widths are smaller than those of the heavy mesons allowed to decay with 
an emission of a strange particle by four orders of magnitude.

\begin{center}
\label{tab:tab_res3}
\tablecaption{Numerical evaluation of the decay widths of excited $B$ mesons}
\tablehead{\hline
Initial State & Final & $k_R$ & $\Gamma_{th}$ 
& $\Gamma_{exp}$ \\
$\left({}^{2S+1}L_J\right)$ & State & (MeV/$c$) & (MeV) 
& (MeV) \\ \hline
}
\begin{supertabular}{|c|c|c|c|c|}[4]
$B_{0}^{*}(5590)^0$ & $B^{\pm}\pi^{\mp}$ & 269.9 & 
$35$ & \\ 
$({}^3P_0)$ & $B^0\pi^0$ & 271.8 & 
$17$ & \\ 
not observed & $B\pi$(sum) &  & 87 & no exp\\ \hline
\hline
$B_{1}(5646)^0$ & $B^{*\pm}\pi^{\mp}$ & 280.7 & 
$37$ & \\ 
$("{}^3P_1")$ & $B^{*0}\pi^0$ & 282.8 & 
$19$ & \\ 
not observed & $B^*\pi$(sum) &  & 93 & no exp\\ \hline
\hline
$B_1(5721)^0$ & $B^{*\pm}\pi^\mp$ & 360.0 & 
7.5 & \\ 
$("{}^1P_1")$ & $B^{*0}\pi^0$ & 361.7 & 
3.8 & \\ 
 & $B^*\pi$(sum) &  & 19 & \\ 
& all & & & no exp\\ \hline
\hline
 & $B^{\pm}\pi^\mp$ & 424.4 & 
6.4 & \\ 
$B_2^{*}(5747)^0$ & $B^{0}\pi^0$ & 425.5 & 
3.2 & \\ 
$({}^3P_2)$ & $B^{*\pm}\pi^\mp$ & 379.5 & 
5.7 & \\ 
 & $B^{*0}\pi^0$ & 381.1 & 
2.9 & \\ 
 & $B\pi+B^*\pi$(sum) &  & 30 & \\ 
& all & & & $22.7\begin{array}{cc}+3.8&+3.2\\
-3.2&-10.2\end{array}$\\ \hline
\hline
 & $B^{\pm}\pi^\mp$ & 651.0 & 
5.8 & \\ 
$B(5985)^0$ & $B^{0}\pi^0$ & 651.6 & 
2.9 & \\ 
$({}^3D_1)$ & $B^{*\pm}\pi^\mp$ & 609.4 & 
2.3 & \\ 
not observed & $B^{*0}\pi^0$ & 610.3 & 
1.1 & \\ 
 & $B\pi+B^*\pi$(sum) &  & 20 & no exp\\
\hline
\end{supertabular}
\end{center}

${}^3P_0\left(0^+\right)$, $"{}^3P_1"\left(1^+\right)$ and ${}^3D_1\left(1^+\right)$ states
of the bottomed 
meson are not yet established, and their masses are taken from our 
paper~\cite{mass_spectrum}.  The only observed decay width of $B_2^*(5747)$ is 
consistent with our prediction.

\begin{center}
\label{tab:tab_res4}
\tablecaption{Numerical evaluation of the decay widths of excited $B_s$ mesons}
\tablehead{\hline
Initial State & Final & $k_R$ & $\Gamma_{th}$ 
& $\Gamma_{exp}$ \\
$\left({}^{2S+1}L_J\right)$ & State & (MeV/$c$) & (MeV) 
& (MeV) \\ \hline
}
\begin{supertabular}{|c|c|c|c|c|}
$B_{s0}^{*}(5615)^0$ & $B_s^0\pi^0$ & 204.2 & 
$1.6\times10^{-3}$ & no exp\\ 
$({}^3P_0)$ & $B_s^{*0}\pi^0$ & - & 
0 & \\ 
not observed & & &
 & \\ \hline
\hline
$B_{s1}(5679)^0$ & $B_s^0\pi^0$ & - & 
0 & \\ 
$("{}^3P_1")$ & $B_s^{*0}\pi^0$ & 224.0 & 
$1.9\times10^{-3}$ & no exp\\ 
not observed &  &  & 
 & \\ \hline
\hline
 & $B^{\pm}K^\mp$ & 229.2 & 
0 & \\ 
$B_{s1}(5830)^0$ & $B^0K^0$ & 230.5 & 
0 & \\ 
$("{}^1P_1")$ & $B^{*\pm}K^\mp$ & 93.6 & 
0.011 & \\ 
 & $B^{*0}K^0$ & 98.3 & 
0.014 & \\ 
 & $B^*K$(sum) & & 0.036 & \\ 
& all & & & no exp\\ \hline
\hline
 & $B^\pm K^\mp$ & 250.6 & 
0.56 & \\ 
$B_{s2}^{*}(5840)^0$ & $B^0K^0$ & 251.8 & 
0.58 & \\ 
$({}^3P_2)$ & $B^{*\pm}K^\mp$ & 135.3 & 
0.041 & \\ 
 & $B^{*0}K^0$ & 138.6 & 
0.046 & \\ 
 & $BK+B^*K$(sum) & & 1.8 & \\ 
& all & & & no exp\\ \hline
\hline
 & $B^\pm K^\mp$ & 523.1 & 
10 & \\ 
$B_{s}(6025)^0$ & $B^0K^0$ & 523.5 & 
10& \\ 
$({}^3D_1)$ & $B^{*\pm}K^\mp$ & 465.9 & 
3.5 & \\ 
not observed & $B^{*0}K^0$ & 466.8 & 
3.5 & \\ 
 & $BK+B^*K$(sum) & & 41 & no exp\\
\hline
\end{supertabular}
\end{center}

${}^3P_0(0^+)$, $"{}^3P_1"(1^+)$ and ${}^3D_1(1^+)$ states of the $B_s$ 
meson are not yet found, and their masses are also taken from our 
paper~\cite{mass_spectrum}.  
As is the $D_s$ meson, the narrow widths of ${}^3P_0(0^+)$ and 
$"{}^3P_1"(1^+)$ 
states of the $B_s$ meson are explained by kinematics.  The predicted mass 
values of these particles are below the the threshold of Konic decay, 
and they decay only through the $\pi^0-\eta$ mixing.

\section{New sum rules on the decay widths in the limit of $m_Q\to\infty$}\label{sec4}
\hspace*{\parindent}
In the limit of $m_Q\to\infty$, the spin of the heavy quark is decoupled from 
the orbital angular momentum and the spin of the light quark, and the radial 
wave functioin is solved for each eigenvalue of $k$. The states with $J^P$ and $(J+1)^P$ take 
the same value of $k$, and they  belong to the same spin multiplet.  They are 
mixed by the spin rotation of the heavy quark, $\mathcal{R}$.  If we denote the 
transformation of the state as
\begin{equation}
\mathcal{R}|\,P,k,p\,>=|\,P,k,q\,>,
\end{equation}
then 
\begin{equation}
\mathcal{R}\left(\sum_{p}|\,P,k,p\,><\,P,k,p\,|\right)\mathcal{R}^\dagger
=\sum_{q}|\,P,k,q\,><\,P,k,q\,|,
\end{equation}
where $P$ stands for the four-mometum of the heavy-light meson, $k$ is 
the eigenvalue to discriminate the spin multiplet, and $p$ and $q$ label the 
polarization of the total spin.  The sum is taken over all states belonging 
to the same spin multiplet.  
Because the axial-vector current for the light quark and
hence the interaction given by Eq.~(\ref{Lint} are invariant under $\mathcal{R}$, 
we obtain the following relation among the transition amplitudes 
\begin{eqnarray}
&& \sum_{p'}|<\,P',k',p'\,|\mathcal{L}_{int}|\,P,k,p\,>|^2\nonumber\\
&=&\sum_{p'}|<\,P',k',p'\,|
\mathcal{R}^\dagger\mathcal{R}\mathcal{L}_{int}\mathcal{R}^\dagger\mathcal{R}|\,P,k,p\,>|^2 
\nonumber\\
&=&<\,P,k,p\,|\mathcal{R}^\dagger\mathcal{L}_{int}\mathcal{R}\left(
\sum_{p'}|\,P',k',p'\,><\,P',k',p'\,|\right)\mathcal{R}^\dagger\mathcal{L}_{int}\mathcal{R}
|\,P,k,p\,>
\nonumber\\
&=&<\,P,k,q\,|\mathcal{L}_{int}
\sum_{q'}|\,P',k',q'\,><\,P',k',q'\,|\mathcal{L}_{int}|\,P,k,q\,>
\nonumber\\
&=&\sum_{q'}|<\,P',k',q'\,|\mathcal{L}_{int}|\,P,k,q\,>|^2.
\end{eqnarray}
Neglecting the mass difference of the mesons in the same multiplet, we 
obtain the following relation on the deccay widths,
\begin{equation}
\sum_{p'}\Gamma\left(\left|\,P,k,p\,\right>\to\left|\,P',k',p'\,\right>
+\pi/K\right)=
\sum_{p'}\Gamma\left(\left|\,P,k,q\,\right>\to\left|\,P',k',p'\,\right>
+\pi/K\right),
\end{equation}
where the sum is taken over all states belonging to the same spin multiplet and
$p$ and $q$ are in the same spin multiplet also.
\begin{eqnarray}
&&\frac{1}{2j+1}\sum_{p(j),p'}\Gamma\left(\left|\,P,k,p(j)\right>\to
\left|\,P',k',p'\,\right>+\pi/K\right)\nonumber\\
&=&
\frac{1}{2(j+1)+1}\sum_{p(j+1),p'}\Gamma\left(\left|\,P,k,p(j+1)\,\right>\to
\left|\,P',k',p'\,\right>+\pi/K\right).
\end{eqnarray}

We are able to explicitly confirm these sum rules in the case of the initial 
states 
$(0^+,1^+)$ with $k=1$ and ~$(1^+,2^+)$ with $k=-2$ and the final states 
$(0^-,1^-)$ with $k=-1$ by using the expressions of the decay widths given in 
Appendix C.

As a matter of fact, there exist mass difference between the mesons belonging 
to the same spin multiplet, and the sum rules are violated more or less by the 
kinematical phase factor.  Among the various decay modes the following sum 
rules are not largely affected by the kinematical phase factor because of relatively
small mass difference between initial and final heavy-light mesons, and these 
rules should be confirmed by experiments,
\begin{eqnarray}
\Gamma\left(D_{s0}^*(0^+,2317)\to D_s(0^-,1968)+\pi\right)&=&\Gamma\left(D_{s1}(1^+,2460)\to D_s^*(1^-,2112)+\pi\right),\nonumber\\
\Gamma\left(B_{s0}^*(0^+,5615)\to B_s(0^-,5366)+\pi\right)&=&\Gamma\left(B_{s1}(1^+,5679)\to B_s^*(1^-,5415)+\pi\right).
\end{eqnarray}
Here the decay modes $0^+\to1^-+\pi$ and $1^+\to0^-+\pi$ are forbidden and 
the sum rules are simply reduced to the above equations.

\section{Conclusions and discussion}\label{sec5}
\hspace*{\parindent}
The wave function of the heavy-light meson in the moving frame is 
approximately related to the one in the rest frame.   Then the the transition 
amplitudes of the excited states of the heavy-light meson to the lower states 
by emitting a chiral particle are expressed 
in terms of the wave functions of the heavy-light meson in the rest frame.  
The plane wave of the emitted particle, $e^{-ikz}$, has been inserted without 
an explanation by all authors in the preceding works, but we have found that 
a phase factor $e^{-2im_QVz}$ should be inserted instead of $e^{-ikz}$.  
The boost operator affects the matrix elements of the components of 
the the axial-vector currents perpendicular to the momentum of the emitted 
chiral particle.  However these componets are irrelevant in our calculation 
because the matrix elements of the axial-vector current are contarcted with 
the momentum of the emitted chiral particle.  

We have used the wave functions of the heavy-light meson obtained by the 
relativistic potential model in Ref.~\cite{mass_spectrum}.  
Beacause ${}^3P_0\left(0^+\right)$, $"{}^3P_1"\left(1^+\right)$ and ${}^3D_1\left(1^+\right)$
states of the $B$ 
and $B_s$ mesons have not yet been observed, we have used our 
predicted mass values for these particles~\cite{mass_spectrum}.  

The partial decay widths of the excited heavy-light meson emitting one chiral 
particle are numerically evaluated, and the predicted values are consistent 
with the experimentally observed total decay widths of excited heavy-light 
mesons.  

One of notable predictions is the narrow decay widths of undiscovered 
$B_s(0^+, 1^+)$ mesons.  Our predicted mass values of these are below the 
threshold of the Konic decay, and they decay through the $\pi^0-\eta$ mixing 
with narrow decay widths of a few keV.  Our predictions are awaited for the experimental confirmation.  

The parameters used in our calculation are not much different from those used by 
Di Pierro and Eichten ~\cite{PE}, but the plain wave of the emitted 
particle($e^{-ikz}$) in the transition amplitudes is replaced by another 
phase factor $e^{-2im_QVz}$ due to the recoil effect.  This replacement makes 
the decay widths large in general, because the oscillatory cancellation due to 
this phase factor becomes weak.

There are 
numerous studies to compute the decay widths using the Schwinger-Dyson 
amplitudes, for example, Ref. \cite{ElBennich:2010ha} and references 
therein.  The coupling constant ${\hat g}$ appearing in these articles and 
our axial-vector coupling constant $g$ appearing in Eq. (\ref{Lint})
are related to each other in the case of $D^*\to D+\pi$ or 
kinematically forbidden case of $B^*\to B+\pi$ as follows:
\begin{eqnarray}
  k^\mu\left<0^-\pi |j_{5\mu}|1^-\right> &=& \left<H\pi|H^*\right>
  \quad (\hbox{in their notation}), \nonumber \\
  \hbox{left hand side} &\approx& \frac{2g\sqrt{M_1M_2}}{f_\pi}
  \left(k^0\eta_{A1}^{(1)}-k^3\eta_{A2}^{(1)}\right)\epsilon_3 \nonumber \\
  &=& -\frac{2g\sqrt{M_1M_2}}{f_\pi} k^3 \eta_{A2}^{(1)} \epsilon_3, \quad
  (\eta_{A1}^{(1)}=0), \nonumber \\
  \hbox{right hand side} &=& -\frac{2\hat{g}\sqrt{M_1M_2}}{f_\pi}k^3\epsilon_3,
  \nonumber \\
  g\eta_{A2}^{(1)} &\approx& \hat{g}.
\end{eqnarray}
$\eta^{(1)}$'s are defined in Eqs. (C1) and (C2) in Appendix C, which are the 
ovelapping integrals of the initial and final wave functions.  
Coupling constants thus determined may be used in the study of 
dissociation processes $\pi+J/\psi\to D+\bar{D}$ or $\pi+\Upsilon\to B+\bar{B}$ 
by exchanging $D^*$ or $B^*$. 
We have obtained $g\eta^{(1)}_{A2}=\sqrt{0.608}\times 0.74=0.577$ while they gave
the value $\hat{g}=0.53$ using the leptonic decay of the heavy-light
system. \cite{ElBennich:2010ha}
Agreement of these values is not surprising because the authors of
Ref. \cite{ElBennich:2010ha} as well as we have fitted couplings with the
experimental data.

In the course of the calculation, new sum rules are found to hold in the limit 
of $m_Q\to\infty$.  Among these rules, 
$
\Gamma(D_{s0}^*(2317)\to D_s+\pi)=\Gamma(D_{s1}(2460)\to D_s^*+\pi)
$ and 
$
\Gamma(B_{s0}^*(5615)\to B_s+\pi)=\Gamma(B_{s1}(5679)\to B_s^*+\pi)
$ are supposed to hold with a good accuracy and should be verified by future 
experiments.

The radiative decays of the heavy-light mesons are under study in the same 
formalism as the present one and the results will be published in near future.

\appendix
\section{Tensor structures of the matrix elements of the axial-vector current}
The matrix elements of the axial-vector current are expressed by
the polarization vectors/tensors in the rest frame of each heavy-light
meson as follows:
\begin{eqnarray}
\frac{<0^-|j_5^\mu|1^->}{2\sqrt{M_2M_1}}&=&
\gamma^2\biggl(\gamma V\varepsilon^3\{\xi^{(k)}_{A1}+2\xi^{(k)}_{A2}\},\,
\varepsilon^1\xi^{(k)}_{A1},\,
\varepsilon^2\xi^{(k)}_{A1},\,
\gamma\varepsilon^3\{\xi^{(k)}_{A1}+2V^2\xi^{(k)}_{A3}\}\biggr),\quad(k=1,10)
\\
\frac{<0^-|j_5^\mu|0^+>}{2i\sqrt{M_2M_1}}&=&
\gamma\left(\xi^{(2)}_{A1},\,
0,\,
0,\,
V\xi^{(2)}_{A2}\right),
\\
\frac{<1^-|j_5^\mu|0^+>}{2\sqrt{M_2M_1}}&=&
\gamma^2V\left(0,\,\varepsilon^{*2},\,-\varepsilon^{*1},\,0\right)\xi^{(3)}_A,
\\
\frac{<0^-|j_5^\mu|1^+>}{2\sqrt{M_2M_1}}  &=&
\gamma^2V\left(0,\,\varepsilon^2,\,-\varepsilon^1,\,0\right)\xi^{(k)}_A
\quad(k=4,5),
\end{eqnarray}
\begin{equation}
\left.
\begin{array}{rl}
\displaystyle{\frac{<1^-|j_5^0|1^+>}{2i\sqrt{M_2M_1}}}&=
-\gamma\left[(\omega\varepsilon^3_1\varepsilon^{*3}_2+\varepsilon^i_1\varepsilon^{*i}_2)\xi^{(k)}_{A1}
+\gamma^2V^2\varepsilon^3_1\varepsilon^{*3}_2\{\xi^{(k)}_{A3}
+\xi^{(k)}_{A4}\}\right],
\\
\displaystyle{\frac{<1^-|j_5^3|1^+>}{2i\sqrt{M_2M_1}}}&=
-\gamma V\left[(\omega\varepsilon^3_1\varepsilon^{*3}_2+\varepsilon^i_1\varepsilon^{*i}_2)\xi^{(k)}_{A2}
+\gamma^2\varepsilon^3_1\varepsilon^{*3}_2\{\xi^{(k)}_{A3}-\xi^{(k)}_{A4}\}\right],
\\
\displaystyle{\frac{<1^-|j_5^1|1^+>}{2i\sqrt{M_2M_1}}}&=
\gamma^2 V\{-\varepsilon^1_1\varepsilon^{*3}_2\xi^{(k)}_{A3}
+\varepsilon^3_1\varepsilon^{*1}_2\xi^{(k)}_{A4}\},
\end{array}\right\}
\quad (k=6,7)
\end{equation}
\begin{eqnarray}
\frac{<0^-|j_5^\mu|2^+>}{2i\sqrt{M_2M_1}} &=&
\gamma^2V\biggl(\gamma V\varepsilon^{33}\left[\xi^{(8)}_{A1}+2\gamma^2\{\xi^{(8)}_{A2}
+\xi^{(8)}_{A3}\}\right],\,
\varepsilon^{13}\xi^{(8)}_{A1},\,
\varepsilon^{23}\xi^{(8)}_{A1},\,\nonumber\\
&& \qquad\qquad
\gamma\varepsilon^{33} \left[\xi^{(8)}_{A1}+2\gamma^2 V^2\{\xi^{(8)}_{A2}
-\xi^{(8)}_{A3}\}\right]\biggr),
\\
\frac{<1^-|j_5^0|2^+>}{2\sqrt{M_2M_1}} &=&
2\gamma^3V^2(\varepsilon^{13}_1\varepsilon^{*2}_2-\varepsilon^{23}_1\varepsilon^{*1})
\{\xi^{(9)}_{A2}-2\gamma^2\xi^{(9)}_{A6}\},
\nonumber\\
\frac{<1^-|j_5^3|2^+>}{2\sqrt{M_2M_1}} &=&
2\gamma^3V(\varepsilon^{13}_1\varepsilon^{*2}_2-\varepsilon^{23}_1\varepsilon^{*1})
\{\xi^{(9)}_{A1}-2\gamma^2V^2\xi^{(9)}_{A7}\},
\nonumber\\
\frac{<1^-|j_5^1|2^+>}{2\sqrt{M_2M_1}} &=&
2\gamma^4V\left\{(\varepsilon^{23}_1\varepsilon^{*3}_2
-\varepsilon^{33}_1\varepsilon^{*2}_2)\xi^{(9)}_{A1}
+V^2(\varepsilon^{23}_1\varepsilon^{*3}_2
+\varepsilon^{33}_1\varepsilon^{*2}_2)\xi^{(9)}_{A2}\right\}\nonumber\\
&& \quad
-\gamma^2V\biggl[(
\omega\varepsilon^{23}_1\varepsilon^{*3}_2+\varepsilon^{2j}_1\varepsilon^{*j}_2)\xi^{(9)}_{A3}
+4\gamma^4V^2\left\{
\varepsilon^{23}_1\varepsilon^{*3}_2\xi^{(9)}_{A4}
-\varepsilon^{33}_1\varepsilon^{*2}_2\xi^{(9)}_{A5}\right\}\biggr],
\\
\frac{<1^-|j_5^0|1^->}{2i\sqrt{M_2M_1}} &=&
\gamma V(\varepsilon^{1}_1\varepsilon^{*2}_2-\varepsilon^{2}_1\varepsilon^{*1}_2)
\{\xi^{(11)}_{A2}-2\gamma^2\xi^{(11)}_{A5}\},
\nonumber\\
\frac{<1^-|j_5^3|1^->}{2i\sqrt{M_2M_1}} &=&
\gamma(\varepsilon^{1}_1\varepsilon^{*2}_2-\varepsilon^{2}_1\varepsilon^{*1}_2)
\{\xi^{(11)}_{A1}-2\gamma^2V^2\xi^{(11)}_{A6}\},
\nonumber\\
\frac{<1^-|j_5^1|1^->}{2i\sqrt{M_2M_1}} &=&
\gamma^2\left\{(\varepsilon^{2}_1\varepsilon^{*3}_2
-\varepsilon^{3}_1\varepsilon^{*2}_2)\xi^{(11)}_{A1}
+V^2(\varepsilon^{2}_1\varepsilon^{*3}_2
+\varepsilon^{3}_1\varepsilon^{*2}_2)\xi^{(11)}_{A2}\right\}\nonumber\\
&& \quad
-2\gamma^4V^2\left\{
\varepsilon^{2}_1\varepsilon^{*3}_2\xi^{(11)}_{A3}
-\varepsilon^{3}_1\varepsilon^{*2}_2\xi^{(11)}_{A4}\right\},
\end{eqnarray}

where the repeated roman indices should be understood as contraction
 with respect to the spatial componets perpendicular to the momentum, 
that is, $\varepsilon^i_1\varepsilon^{*i}_2\equiv\sum_{i=1,2}\varepsilon^i_1\varepsilon^{*i}_2$.  
The lower indices 1 and 2 of $M$ and the polarization vectors/tensors stand
 for the initial and the final one, respectively.

\vskip 12pt

\section{The angular-spin part of the wave functions}
Polarization vectors and tensors of mesons satisfy the following
orthonormal conditions and completeness conditions,
\begin{equation}
({\vec \varepsilon}^{\;(p)*}\cdot{\vec \varepsilon}^{\;(q)})=
\delta_{pq}\quad(p,q=1,2,3),
\end{equation}
\begin{equation}
\sum_{p}\varepsilon^{(p)*}_i\varepsilon^{(p)}_j=\delta_{ij},
\end{equation}
\begin{equation}
\sum_{j,k}\varepsilon^{(p)*}_{jk}\varepsilon^{(q)}_{jk}=\delta_{pq}
\quad(p,q=1,2,3,4,5)
\end{equation}
\begin{equation}
\sum_{p}\varepsilon^{(p)}_{jk}\varepsilon^{(p)*}_{j'k'}=
\frac{1}{2}(\delta_{jj'}\delta_{kk'}+\delta_{jk'}\delta_{kj'}-
\frac{2}{3}\delta_{jk}\delta_{j'k'}),
\end{equation}

The angular-spin part of the wave functions of various spins take the
 expression with the polarization vectors or tensors as listed in Table
~\ref{tab:tab1}. 
\begin{table}[H]
\caption{the angular-spin part of wave functions of low lying states}
\label{tab:tab1}
\begin{center}
\begin{tabular}{|c|c|c|} \hline
${}^{2S+1}L_J$ & $k$ & $ y^{(p)}$\\ \hline\hline
${}^1S_0$ & $-1$ & $1$\\ \hline
${}^3S_1$ & $-1$ & ${\vec \sigma}\cdot{\vec \varepsilon}^{\;(p)}$\\ \hline
${}^3P_0$ & $1$ & ${\vec n}\cdot{\vec \sigma}$\\ \hline
$"{}^3P_1"$ & $1$ & 
$({\vec n}\cdot{\vec \sigma})
({\vec \sigma}\cdot{\vec \varepsilon}^{\;(p)})$\\ \hline
$"{}^1P_1"$ & $-2$ & 
$\frac{1}{\sqrt{2}}\left\{
3{\vec n}\cdot{\vec \varepsilon}^{\;(p)}-
({\vec n}\cdot{\vec \sigma})
({\vec \sigma}\cdot{\vec \varepsilon}^{\;(p)})\right\}$\\ \hline
${}^3P_2$ & $-2$ & 
$\sqrt{3}n_j\sigma_k\varepsilon_{jk}$\\ \hline
${}^3D_1$ & $2$ & 
$\frac{1}{\sqrt{2}}\left\{
3({\vec n}\cdot{\vec \varepsilon}^{\;(p)})({\vec n}\cdot{\vec \sigma})
-({\vec \sigma}\cdot{\vec \varepsilon}^{\;(p)})\right\}$\\ \hline
\end{tabular}
\end{center}
\end{table}

They satisfy the orthonormal conditions,
\begin{equation}
\int\frac{d\Omega}{4\pi}\frac{1}{2}\text{tr}[y^{(p)*}y^{(q)}]=\delta_{pq},
\end{equation}
where $p$ and $q$ represent both the polarization state and the spin state.
\section{The decay widths in terms of the radial wave functions}
The transition amplitudes are expressed by the radial wave functions and 
the spherical Bessel function, $f(x)\equiv \sin{x}/x$, as follows. \\
\noindent
(1)\quad $1^-(^3S_1)\rightarrow0^-(^1S_0)$
\begin{eqnarray}
\eta^{(1)}_{A1}&\equiv& V(\xi^{(1)}_{A1}+2\xi^{(1)}_{A2})
=-\int dr\,f'(qr)(u_2v_1-v_2u_1)+O(V^3)
\\
\eta^{(1)}_{A2}&\equiv&\xi^{(1)}_{A1}+2V^2\xi^{(1)}_{A3}
=\int dr\biggl[f(qr)u_2u_1-\left\{f(qr)+2f''(qr)\right\}v_2v_1\biggr]+O(V^2)
\\
\xi^{(1)}_{A1}&=&\int dr\biggl[f(qr)u_2u_1+f''(qr)v_2v_1
-Vf'(qr)(u_2v_1+v_2u_1)\biggr]+O(V^2)
\\
\Gamma&=&
\frac{(g\zeta)^2k_RM_2}{6\pi M_1f_\pi^2}
\biggl\{\omega_B\eta^{(1)}_{A1}-k_B\eta^{(1)}_{A2}\biggr\}^2
\end{eqnarray}

\noindent
(2)\quad $0^+(^3P_0)\rightarrow0^-(^1S_0)$
\begin{eqnarray}
\xi^{(2)}_{A1}&=&
-\int dr\,f(qr)(u_2v_1-v_2u_1)+O(V^2)
\\
\eta^{(2)}_A&\equiv& V\xi^{(2)}_{A2}
=-\int dr\,f'(qr)(u_2u_1+v_2v_1)+O(V^3)
\\
\Gamma&=&\frac{(g\zeta)^2M_2k_R}{2\pi f_\pi^2 M_1}
\biggl\{\omega_B\xi^{(2)}_{A1}-k_B\eta^{(2)}_A\biggr\}^2
\end{eqnarray}

\noindent
(3)\quad $0^+(^3P_0)\rightarrow1^-(^3S_1)$
\begin{eqnarray}
V\xi^{(3)}_A&=&\int dr\biggl\{-f'(qr)(u_2u_1-v_2v_1)
-Vf(qr)u_2v_1+Vf''(qr)v_2u_1\biggr\}+O(V^3)
\\
\Gamma&=&0
\end{eqnarray}

\noindent
(4)\quad $1^+("^3P_1")\rightarrow0^-(^1S_0)$
\begin{eqnarray}
V\xi^{(4)}_A&=&\int dr\biggl\{-f'(qr)(u_2u_1-v_2v_1)
-Vf(qr)u_2v_1+Vf''(qr)v_2u_1\biggr\}+O(V^3)
\\
\Gamma&=&0
\end{eqnarray}

\noindent
(5)\quad $1^+("^1P_1")\rightarrow0^-(^1S_0)$
\begin{eqnarray}
V\xi^{(5)}_A&=&\frac{1}{\sqrt{2}}\int dr\biggl[f'(qr)(u_2u_1-v_2v_1)
-\frac{V}{2}\left\{f(qr)+3f''(qr)\right\}u_2v_1\nonumber\\&&\qquad
+\frac{V}{2}\left\{3f(qr)+f''(qr)\right\}v_2u_1\biggr]+O(V^3)
\\
\Gamma&=&0
\end{eqnarray}

\noindent
(6)\quad $1^+("^3P_1")\rightarrow1^-(^3S_1)$
\begin{eqnarray}
\xi^{(6)}_{A1}
&=&\int dr\,f(qr)(u_2v_1-v_2u_1)+O(V^2)
\\
\eta^{(6)}_A&\equiv& V\xi^{(6)}_{A2}
=\int dr\,f'(qr)(u_2u_1+v_2v_1)+O(V^3)
\\
V\xi^{(6)}_{A3}
&=&V\xi^{(6)}_{A4}
=\int dr\biggl\{-f'(qr)(u_2u_1-v_2v_1)
-Vf(qr)u_2v_1
+Vf''(qr)v_2u_1\biggr\}+O(V^3)
\\
\Gamma&=&
\frac{(g\zeta)^2k_RM_2}{2\pi M_1f_\pi^2}
\biggl\{\omega_B\xi^{(6)}_{A1}-k_B\eta^{(6)}_A\biggr\}^2
\end{eqnarray}

\noindent
(7)\quad $1^+("^1P_1")\rightarrow1^-(^3S_1)$
\begin{eqnarray}
\xi^{(7)}_{A1}
&=&\frac{1}{2\sqrt{2}}
\int dr\left\{f(qr)+3f''(qr)\right\}(u_2v_1-v_2u_1)+O(V^4)
\\
\eta^{(7)}_{A1}&\equiv&
\xi^{(7)}_{A1}+V^2\left\{\xi^{(7)}_{A3}+\xi^{(7)}_{A4}\right\}
=-2\xi^{(7)}_{A1}+O(V^4)
\\
\eta^{(7)}_{A2}&\equiv& V\xi^{(7)}_{A2}
=\frac{1}{\sqrt{2}}\int dr\biggl[-f'(qr)u_2u_1
+\left\{2f'(qr)+3f'''(qr)\right\}v_2v_1\biggr]+O(V^3)
\\
\eta^{(7)}_{A3}&\equiv& V\left\{
\xi^{(7)}_{A2}+\xi^{(7)}_{A3}-\xi^{(7)}_{A4}\right\}
=-2\eta^{(7)}_{A2}+O(V^3)
\\
V\xi^{(7)}_{A3}
&=&\frac{1}{\sqrt{2}}
\int dr\biggl[f'(qr)u_2u_1+\left\{2f'(qr)+3f'''(qr)\right\}v_2v_1
\nonumber\\&&\qquad
-\frac{V}{2}\left\{f(qr)+3f''(qr)\right\}u_2v_1
-\frac{V}{2}\left\{3f(qr)+5f''(qr)\right\}v_2u_1\biggr]+O(V^3)
\\
V\xi^{(7)}_{A4}
&=&\frac{1}{\sqrt{2}}
\int dr\biggl[-2f'(qr)u_2u_1-\left\{f'(qr)+3f'''(qr)\right\}v_2v_1
\nonumber\\&&\qquad
+V\left\{f(qr)+3f''(qr)\right\}u_2v_1
+2Vf''(qr)v_2u_1\biggr]+O(V^3)
\\
\Gamma
&=&\frac{(g\zeta)^2k_RM_2}{2\pi M_1f_\pi^2}\frac{1}{3}\biggl[
\left\{\omega_B\xi^{(7)}_{A1}-k_B\eta^{(7)}_{A2}\right\}^2\times2
+\left\{\omega_B\eta^{(7)}_{A1}-k_B\eta^{(7)}_{A3}\right\}^2
\biggr]
\nonumber\\
&=&\frac{(g\zeta)^2k_RM_2}{\pi M_1f_\pi^2}
\biggl\{\omega_B\xi^{(7)}_{A1}-k_B\eta^{(7)}_{A2}\biggr\}^2
\end{eqnarray}

\noindent
(8)\quad $2^+(^3P_2)\rightarrow0^-(^1S_0)$
\begin{eqnarray}
\eta^{(8)}_{A1}&\equiv& V^2\left\{
\xi^{(8)}_{A1}+2\xi^{(6)}_{A2}+2\xi^{(6)}_{A3}\right\}
\nonumber\\
&=&\frac{\sqrt{3}}{2}
\int dr\left\{f(qr)+3f''(qr)\right\}(u_2v_1-v_2u_1)+O(V^4)
\\
\eta^{(8)}_{A2}&\equiv& V\biggl[
\xi^{(8)}_{A1}+2V^2\left\{\xi^{(6)}_{A2}-\xi^{(6)}_{A3}\right\}\biggr]
\nonumber\\
&=&\sqrt{3}\int dr\biggl[-f'(qr)u_2u_1
+\left\{2f'(qr)+3f'''(qr)\right\}v_2v_1\biggr]+O(V^3)
\\
V\xi^{(8)}_{A1}
&=&\sqrt{3}\int dr\biggl[-f'(qr)u_2u_1-\left\{f'(qr)+2f'''(qr)\right\}v_2v_1
\nonumber\\&&\qquad
+\frac{V}{2}\left\{f(qr)+3f''(qr)\right\}(u_2v_1+v_2u_1)\biggr]+O(V^3)
\\
\Gamma
&=&\frac{(g\zeta)^2k_RM_2}{15\pi M_1f_\pi^2}
\biggl\{\omega_B\eta^{(8)}_{A1}-k_B\eta^{(8)}_{A2}\biggr\}^2
\end{eqnarray}

\noindent
(9)\quad $2^+(^3P_2)\rightarrow1^-(^3S_1)$
\begin{eqnarray}
\eta^{(9)}_{A1}&\equiv& 2V^2\left\{
\xi^{(9)}_{A2}-2\xi^{(9)}_{A6}\right\}
\nonumber\\
&=&-\frac{\sqrt{3}}{2}
\int dr\left\{f(qr)+3f''(qr)\right\}(u_2v_1-v_2u_1)+O(V^4)
\\
\eta^{(9)}_{A2}&\equiv& 2V\left\{
\xi^{(9)}_{A1}-2V^2\xi^{(9)}_{A7}\right\}
\nonumber\\
&=&\sqrt{3}\int dr\biggl[f'(qr)u_2u_1
-\left\{2f'(qr)+3f'''(qr)\right\}v_2v_1\biggr]+O(V^3)
\\
-V\xi^{(9)}_{A3}
&=&\sqrt{3}\int dr\biggl[-\left\{f'(qr)+f'''(qr)\right\}v_2v_1
\nonumber\\&&\qquad
+V\left\{f(qr)+f''(qr)\right\}v_2u_1\biggr]+O(V^3)
\\
\eta^{(9)}_{A3}&\equiv& 2V\biggl[
-\xi^{(9)}_{A1}+V^2\left\{\xi^{(9)}_{A2}+2\xi^{(9)}_{A5}\right\}\biggr]
\nonumber\\
&=&\sqrt{3}\int dr\biggl[-f'(qr)u_2u_1-\left\{f'(qr)+2f'''(qr)\right\}v_2v_1
\nonumber\\&&\qquad
+\frac{V}{2}\left\{f(qr)+3f''(qr)\right\}(u_2v_1+v_2u_1)\biggr]+O(V^3)
\\
\Gamma
&=&\frac{(g\zeta)^2k_RM_2}{10\pi M_1f_\pi^2}
\biggl\{\omega_B\eta^{(9)}_{A1}-k_B\eta^{(9)}_{A2}\biggr\}^2
\end{eqnarray}

\noindent
(10)\quad $1^-(^3D_1)\rightarrow0^-(^1S_0)$
\begin{eqnarray}
\eta^{(10)}_{A1}&\equiv& V(\xi^{(10)}_{A1}+2\xi^{(10)}_{A2})
=-\sqrt{2}\int dr\,f'(qr)(u_2v_1-v_2u_1)+O(V^3)
\\
\eta^{(10)}_{A2}&\equiv&\xi^{(10)}_{A1}+2V^2\xi^{(10)}_{A3}\nonumber\\
&=&\frac{1}{\sqrt{2}}\int dr\biggl[-\left\{f(qr)+3f''(qr)\right\}u_2u_1
+\left\{f(qr)-f''(qr)\right\}v_2v_1\biggr]+O(V^2)
\\
\Gamma&=&
\frac{(g\zeta)^2k_RM_2}{6\pi M_1f_\pi^2}
\biggl\{\omega_B\eta^{(10)}_{A1}-k_B\eta^{(10)}_{A2}\biggr\}^2
\end{eqnarray}

\noindent
(11)\quad $1^-(^3D_1)\rightarrow1^-(^3S_1)$
\begin{eqnarray}
\eta^{(11)}_{A1}&\equiv& V(\xi^{(11)}_{A2}+2\xi^{(11)}_{A5})
=\frac{1}{\sqrt{2}}\int dr\,f'(qr)(u_2v_1-v_2u_1)+O(V^3)
\\
\eta^{(11)}_{A2}&\equiv&\xi^{(11)}_{A1}+2V^2\xi^{(11)}_{A6}\nonumber\\
&=&\frac{1}{2\sqrt{2}}\int dr\biggl[\left\{f(qr)+3f''(qr)\right\}u_2u_1
+\left\{-f(qr)+f''(qr)\right\}v_2v_1\biggr]+O(V^2)
\\
\Gamma&=&
\frac{(g\zeta)^2k_RM_2}{3\pi M_1f_\pi^2}
\biggl\{\omega_B\eta^{(11)}_{A1}-k_B\eta^{(11)}_{A2}\biggr\}^2
\end{eqnarray}
%
As defined in Section II, $q=-2m_QV$.  $V$ is the velocity of the heavy-light 
meson in the Breit frame.  $\omega_B$ and $k_B$ denote the energy and 
momentum of the emitted chiral particle in the Breit frame, respectively.
$\zeta$, which appears in the final expression of $\Gamma$, is
$\frac{1}{\sqrt{2}}$ times the coefficient in front of the chiral field in
the matrix Eq.~(\ref{eq:octet}).


\end{document}